\documentclass[10pt]{article}

\begin{document}

\author{C. Bizdadea\thanks{%
E-mail: bizdadea@central.ucv.ro} and S. O. Saliu\thanks{%
E-mail: osaliu@central.ucv.ro}\\
Faculty of Physics, University of Craiova\\
13 A. I. Cuza Str., Craiova RO-1100, Romania}
\title{Lagrangian Sp(3) BRST symmetry for Yang-Mills theory}

\maketitle

\begin{abstract}
The Sp(3) BRST symmetry for Yang-Mills theory is derived in the framework of
the antibracket-antifield formalism.

\end{abstract}

The BRST theory~\cite{1,2,9,8,12} has evolved until being regarded as an
extremely powerful assembly of rules allowing for a systematical quantum
treatment of gauge theories. In the meantime, it was observed that the
analysis of theories subject to gauge invariances, like renormalizability,
anomalies and structure of non-minimal sectors, might be clarified via the
introduction of extended BRST symmetries, like, for instance, the Sp(2)
symmetries~\cite{20,22,29,37}. Recently, it appeared a growing interest for
constructing even more extended symmetries, namely, the Sp(3) BRST symmetry,
at both the Hamiltonian~\cite{38} and Lagrangian~\cite{sp3gen} levels.

In this paper we construct the Lagrangian Sp(3) BRST symmetry for 
Yang-Mills theories. For proofs and details we refer to the
general results from~\cite{sp3gen}. Our procedure goes as follows.
First, we triplicate the gauge transformations of
Yang-Mills fields and derive the ghost and antifield spectra. Second, we
solve the fundamental equation of the Lagrangian Sp(3) BRST formalism,
called the extended classical master equation. Third, we implement a
gauge-fixing procedure that ensures a direct link with the standard
antibracket-antifield approach, and consequently obtain the gauge-fixed
action. Finally, we determine the Sp(3) BRST symmetry of the resulting
gauge-fixed action.

We begin with the action of pure Yang-Mills theory 
\begin{equation}
\label{1}
S_{0}^{\mathrm{L}}\left[ A_{\alpha }^{a}\right] =-\frac{1}{4}\int
d^{D}x\,F_{\alpha \beta }^{a}F_{a}^{\alpha \beta }\,,
\end{equation}
where the field strength of the Yang-Mills fields is defined in the usual
manner by $F_{\alpha \beta }^{a}=\partial _{\alpha }A_{\beta }^{a}-\partial
_{\beta }A_{\alpha }^{a}-f_{bc}^{a}A_{\alpha }^{b}A_{\beta }^{c}$. 
Action~(\ref{1}) is invariant under the gauge transformations 
\begin{equation}
\label{2}
\delta _{\epsilon }A_{\alpha }^{a}=\left( D_{\alpha }\right)
_{\;\;b}^{a}\epsilon ^{b}\,,
\end{equation}
with the covariant derivative expressed by $\left( D_{\alpha }\right)
_{\;\;b}^{a}=\delta _{\;\;b}^{a}\partial _{\alpha }+f_{bc}^{a}A_{\alpha
}^{c} $.

In order to construct a Lagrangian Sp(3) BRST symmetry for this model we
triplicate both the gauge generators and gauge parameters, and work, instead
of~(\ref{2}), with the modified invariances 
\begin{equation}
\label{3}
\delta _{\epsilon }A_{\alpha }^{a}=\left( 
\begin{array}{lll}
\left( D_{\alpha }\right) _{\;\;b}^{a} & \left( D_{\alpha }\right)
_{\;\;b}^{a} & \left( D_{\alpha }\right) _{\;\;b}^{a}
\end{array}
\right) \left( 
\begin{array}{l}
\epsilon _{1}^{b} \\ 
\epsilon _{2}^{b} \\ 
\epsilon _{3}^{b}
\end{array}
\right) \,,
\end{equation}
which are found off-shell second-stage reducible, with the first-,
respectively, second-stage reducibility functions given by 
\begin{equation}
\label{4}
Z_{\;\;C}^{B}=\left( 
\begin{array}{ccc}
\mathbf{0} & \delta _{\;\;c}^{b} & -\delta _{\;\;c}^{b} \\ 
-\delta _{\;\;c}^{b} & \mathbf{0} & \delta _{\;\;c}^{b} \\ 
\delta _{\;\;c}^{b} & -\delta _{\;\;c}^{b} & \mathbf{0}
\end{array}
\right) \,, Z_{\;\;d}^{C}=\left( 
\begin{array}{c}
-\delta _{\;\;d}^{c} \\ 
-\delta _{\;\;d}^{c} \\ 
-\delta _{\;\;d}^{c}
\end{array}
\right) \,.
\end{equation}
According to the general results from~\cite{sp3gen}, we can construct a
Lagrangian Sp(3) BRST symmetry, and hence a BRST tricomplex generated by
three anticommuting differentials $\left( s_{m}\right) _{m=1,2,3}$ 
\begin{equation}
\label{5}
s_{m}s_{n}+s_{n}s_{m}=0\,, m,n=1,2,3\,,
\end{equation}
that start like 
\begin{equation}
\label{6}
s_{m}=\delta _{m}+D_{m}+\cdots \,,
\end{equation}
where $\left( \delta _{m}\right) _{m=1,2,3}$ are the three differentials
from the Koszul-Tate tricomplex, that furnish a triresolution of smooth
functions defined on the stationary surface of field equations, and $\left(
D_{m}\right) _{m=1,2,3}$ represent the exterior derivatives along the gauge
orbits associated with the new second-stage redundant description of the
gauge orbits due to the triplication of the gauge transformations like 
in~(\ref{3}).

The trigrading of the Sp(3) BRST algebra is governed by the ghost tridegree (%
$\mathrm{trigh}=\left( \mathrm{gh}_{1},\mathrm{gh}_{2},\mathrm{gh}%
_{3}\right) $), and we have that $\mathrm{trigh}\left( s_{1}\right) =\left(
1,0,0\right) $, $\mathrm{trigh}\left( s_{2}\right) =\left( 0,1,0\right) $,
and $\mathrm{trigh}\left( s_{3}\right) =\left( 0,0,1\right) $. The ghost
spectrum contains, due to the triplication, the fields 
\begin{equation}
\label{8}
\stackrel{(1,0,0)}{\eta }_{1}^{a},\stackrel{(0,1,0)}{\eta }_{2}^{a},%
\stackrel{(0,0,1)}{\eta }_{3}^{a},\stackrel{(0,1,1)}{\pi }_{1}^{a},\stackrel{%
(1,0,1)}{\pi }_{2}^{a},\stackrel{(1,1,0)}{\pi }_{3}^{a},\stackrel{(1,1,1)}{%
\lambda }^{a},
\end{equation}
where we denoted an element $F$ with the ghost tridegree equal to $\left(
i,j,k\right) $ by $\stackrel{(i,j,k)}{F}$, and set $\mathrm{trigh}\left(
A_{\mu }^{a}\right) =\left( 0,0,0\right) $. The Grassmann parities of the
ghosts are: $\varepsilon \left( \eta _{m}^{a}\right) =\varepsilon \left(
\lambda ^{a}\right) =1$, $\varepsilon \left( \pi _{m}^{a}\right) =0$, $%
m=1,2,3$. The ghost spectrum can be understood by using the properties of
the exterior derivatives along the gauge orbits~\cite{sp3gen}.

The Sp(3) formalism relies on the presence of three antibrackets, denoted by 
$\left( ,\right) _{m}$, $m=1,2,3$, which implies that we have to introduce
three antifields conjugated to each field/ghost, one for each antibracket.
Consequently, the antifield spectrum reads as 
\begin{equation}
\label{9}
\stackrel{(-1,0,0)}{A}_{a}^{*(1)\alpha },\stackrel{(0,-1,0)}{A}%
_{a}^{*(2)\alpha },\stackrel{(0,0,-1)}{A}_{a}^{*(3)\alpha },\stackrel{%
(-2,0,0)}{\eta }_{1a}^{*(1)},\stackrel{(-1,-1,0)}{\eta }_{1a}^{*(2)},%
\stackrel{(-1,0,-1)}{\eta }_{1a}^{*(3)},
\end{equation}
\begin{equation}
\label{10}
\stackrel{(-1,-1,0)}{\eta }_{2a}^{*(1)},\stackrel{(0,-2,0)}{\eta }%
_{2a}^{*(2)},\stackrel{(0,-1,-1)}{\eta }_{2a}^{*(3)},\stackrel{(-1,0,-1)}{%
\eta }_{3a}^{*(1)},\stackrel{(0,-1,-1)}{\eta }_{3a}^{*(2)},\stackrel{(0,0,-2)%
}{\eta }_{3a}^{*(3)},
\end{equation}
\begin{equation}
\label{11}
\stackrel{(-1,-1,-1)}{\pi }_{1a}^{*(1)},\stackrel{(0,-2,-1)}{\pi }%
_{1a}^{*(2)},\stackrel{(0,-1,-2)}{\pi }_{1a}^{*(3)},\stackrel{(-2,0,-1)}{\pi 
}_{2a}^{*(1)},\stackrel{(-1,-1,-1)}{\pi }_{2a}^{*(2)},\stackrel{(-1,0,-2)}{%
\pi }_{2a}^{*(3)},
\end{equation}
\begin{equation}
\label{12}
\stackrel{(-2,-1,0)}{\pi }_{3a}^{*(1)},\stackrel{(-1,-2,0)}{\pi }%
_{3a}^{*(2)},\stackrel{(-1,-1,-1)}{\pi }_{3a}^{*(3)},\stackrel{(-2,-1,-1)}{%
\lambda }_{a}^{*(1)},\stackrel{(-1,-2,-1)}{\lambda }_{a}^{*(2)},\stackrel{%
(-1,-1,-2)}{\lambda }_{a}^{*(3)}\,.
\end{equation}
In order to ensure the nilpotency, as
well as the acyclicity of the Koszul-Tate differentials, we further enlarge
the antifield spectrum with the bar and tilde variables~\cite{sp3gen} 
\begin{equation}
\label{13}
\stackrel{(0,-1,-1)}{\bar{A}}_{a}^{(1)\alpha },\stackrel{(-1,0,-1)}{\bar{A}}%
_{a}^{(2)\alpha },\stackrel{(-1,-1,0)}{\bar{A}}_{a}^{(3)\alpha },\stackrel{%
(-1,-1,-1)}{\bar{\eta}}_{1a}^{(1)},\stackrel{(-2,0,-1)}{\bar{\eta}}%
_{1a}^{(2)},\stackrel{(-2,-1,0)}{\bar{\eta}}_{1a}^{(3)},
\end{equation}
\begin{equation}
\label{14}
\stackrel{(0,-2,-1)}{\bar{\eta}}_{2a}^{(1)},\stackrel{(-1,-1,-1)}{\bar{\eta}}%
_{2a}^{(2)},\stackrel{(-1,-2,0)}{\bar{\eta}}_{2a}^{(3)},\stackrel{(0,-1,-2)}{%
\bar{\eta}}_{3a}^{(1)},\stackrel{(-1,0,-2)}{\bar{\eta}}_{3a}^{(2)},\stackrel{%
(-1,-1,-1)}{\bar{\eta}}_{3a}^{(3)},
\end{equation}
\begin{equation}
\label{15}
\stackrel{(0,-2,-2)}{\bar{\pi}}_{1a}^{(1)},\stackrel{(-1,-1,-2)}{\bar{\pi}}%
_{1a}^{(2)},\stackrel{(-1,-2,-1)}{\bar{\pi}}_{1a}^{(3)},\stackrel{(-1,-1,-2)%
}{\bar{\pi}}_{2a}^{(1)},\stackrel{(-2,0,-2)}{\bar{\pi}}_{2a}^{(2)},\stackrel{%
(-2,-1,-1)}{\bar{\pi}}_{2a}^{(3)},
\end{equation}
\begin{equation}
\label{16}
\stackrel{(-1,-2,-1)}{\bar{\pi}}_{3a}^{(1)},\stackrel{(-2,-1,-1)}{\bar{\pi}}%
_{3a}^{(2)},\stackrel{(-2,-2,0)}{\bar{\pi}}_{3a}^{(3)},\stackrel{(-1,-2,-2)}{%
\bar{\lambda}}_{a}^{(1)},\stackrel{(-2,-1,-2)}{\bar{\lambda}}_{a}^{(2)},%
\stackrel{(-2,-2,-1)}{\bar{\lambda}}_{a}^{(3)},
\end{equation}
\begin{equation}
\label{17}
\stackrel{(-1,-1,-1)}{\tilde{A}}_{a}^{\alpha },\stackrel{(-2,-1,-1)}{\tilde{%
\eta}}_{1a},\stackrel{(-1,-2,-1)}{\tilde{\eta}}_{2a},\stackrel{(-1,-1,-2)}{%
\tilde{\eta}}_{3a},
\end{equation}
\begin{equation}
\label{18}
\stackrel{(-1,-2,-2)}{\tilde{\pi}}_{1a},\stackrel{(-2,-1,-2)}{\tilde{\pi}}%
_{2a},\stackrel{(-2,-2,-1)}{\tilde{\pi}}_{3a},\stackrel{(-2,-2,-2)}{\tilde{%
\lambda}}_{a}\,.
\end{equation}
The Grassmann parities of the antifields~(\ref{9}--\ref{12}) and~(\ref{17}--%
\ref{18}) are opposite to those of the corresponding fields/ghosts, while
those of~(\ref{13}--\ref{16}) coincide with them. The antifields~(\ref{9}--%
\ref{12}) bear a supplementary superscript between parentheses, which
signifies in what bracket they are conjugated to the corresponding
fields/ghosts. The antifield
sector is also graded by a supplementary tridegree, named resolution
tridegree, and defined by $\mathrm{trires}=\left( \mathrm{res}_{1},\mathrm{%
res}_{2},\mathrm{res}_{3}\right) =-\mathrm{trigh}$. The induced simple
grading, called total resolution degree, $\mathrm{res}=\mathrm{res}_{1}+%
\mathrm{res}_{2}+\mathrm{res}_{3}$, will be useful in the sequel when
solving the fundamental equation of the Sp(3) formalism, namely, the
extended classical master equation.

With the ghost and antifield spectra at hand, we can give the boundary
conditions on the solution to the extended classical master equation. They
take the form 
\begin{equation}
\label{sp3.71}
\stackrel{\lbrack 0]}{S}=S_{0}^{\mathrm{L}}\left[ A_{\alpha }^{a}\right] \,,
 \stackrel{[1]}{S}=
\int d^{D}x\,A_{a}^{*(m)\alpha }\left( D_{\alpha }\right)
_{\;\;b}^{a}\eta _{m}^{b}\,,
\end{equation}
\begin{equation}
\label{sp3.72}
\stackrel{\lbrack 2]}{S}=\int d^{D}x\left( \left( \varepsilon _{mnp}\eta
_{na}^{*(m)}+\bar{A}_{b}^{(p)\alpha }\left( D_{\alpha }\right)
_{\;\;a}^{b}\right) \pi _{p}^{a}+\cdots \right) \,,
\end{equation}
\begin{equation}
\label{sp3.73}
\stackrel{\lbrack 3]}{S}=\int d^{D}x\left( -\left( \pi _{ma}^{*(m)}+\bar{\eta%
}_{ma}^{(m)}-\tilde{A}_{b}^{\alpha }\left( D_{\alpha }\right)
_{\;\;a}^{b}\right) \lambda ^{a}+\cdots \right) \,,
\end{equation}
where the supplementary superscript between brackets
in $\stackrel{\lbrack 0]}{S}$, $\stackrel{[1]}{S}$, etc., refers to a
decomposition of the solution to the master equation via the total
resolution degree, and $\varepsilon _{mnp}$ is completely antisymmetric,
with $\varepsilon _{123}=+1$. The boundary conditions~(\ref{sp3.71}--\ref
{sp3.73}) can be understood by means of homological arguments~\cite{sp3gen}.

The construction of the Sp(3) algebra~(\ref{5}) is completely equivalent to
the construction of the anticanonical generator $S$ of this symmetry, which
is solution to the extended classical master equation 
\begin{equation}
\label{7}
\frac{1}{2}\left( S,S\right) +VS=0\,,
\end{equation}
and is required to have the ghost tridegree $\mathrm{trigh}\left( S\right)
=\left( 0,0,0\right) $. The symbol $\left( ,\right) $ denotes the total
antibracket, written as the sum among the three antibrackets, and $V$
represents the noncanonical part of the total Koszul-Tate differential, $%
\delta =\delta _{1}+\delta _{2}+\delta _{3}$.
The master equation~(\ref{7})
projected on independent components reads as $\frac{1}{2}\left( S,S\right)
_{m}+V_{m}S=0$, with $\left( ,\right) _{m}$ and $V_{m}$ obviously meaning
the antibracket, respectively, the noncanonical part of the total
Koszul-Tate differential corresponding to the component `$m$' of the Sp(3)
BRST symmetry, $\delta _{m}=\delta _{\mathrm{can}m}+V_{m}$. 
We mention that the individual antibrackets $\left( ,\right) _{m}$, as
well as the total one, satisfy the usual properties of the 
antibracket-antifield formalism, while the operators $V_{m}$ and $V$
behave like derivations with respect to the antibrackets. Their features
are $\mathrm{trigh}\left( \left( ,\right) _{m} \right) =
\mathrm{trigh}\left( s_{m}\right) $, $\mathrm{trigh}\left( V_{m}\right) =
\mathrm{trigh}\left( s_{m}\right) =-\mathrm{trires}\left( V_{m}\right) $.
The operators $%
V_{m}$ act only on the bar and tilde variables through 
\begin{equation}
\label{20}
V_{m}\bar{\Phi}_{A}^{(n)}=\left( -\right) ^{\varepsilon _{A}}\varepsilon
_{mnp}\Phi _{A}^{*(p)}\,, V_{m}\tilde{\Phi}_{A}=
\left( -\right) ^{\varepsilon
_{A}+1}\bar{\Phi}_{A}^{(m)}\,,
\end{equation}
where $\varepsilon _{A}$ stands for the Grassmann parity of a generic
field/ghost $\Phi ^{A}$.

In order to solve the equation~(\ref{7}), we develop $S$ according to the
total resolution degree, $S=\sum\nolimits_{k\geq 0}\stackrel{[k]}{S}$, $%
\mathrm{res}\left( \stackrel{[k]}{S}\right) =k$, $\mathrm{trigh}\left( 
\stackrel{[k]}{S}\right) =\left( 0,0,0\right) $, where the boundary terms
are given in~(\ref{sp3.71}--\ref{sp3.73}). Following this line, we find that
the solution to~(\ref{7}) reads as 
\begin{eqnarray}
\label{21}
&&S=\int d^{D}x\left( -\frac{1}{4}F_{\alpha \beta }^{a}F_{a}^{\alpha \beta
}+A_{a}^{*(m)\alpha }\left( D_{\alpha }\right) _{\;\;b}^{a}\eta
_{m}^{b}+\left( \varepsilon _{mnp}\eta _{na}^{*(m)}+\bar{A}_{b}^{(p)\alpha
}\left( D_{\alpha }\right) _{\;\;a}^{b}\right) \pi _{p}^{a}+\right.  
\nonumber \\
&&\left( -\pi _{ma}^{*(m)}-\bar{\eta}_{ma}^{(m)}+\tilde{A}_{b}^{\alpha
}\left( D_{\alpha }\right) _{\;\;a}^{b}\right) \lambda ^{a}+\frac{1}{2}\eta
_{na}^{*(m)}f_{bc}^{a}\eta _{m}^{c}\eta _{n}^{b}-\frac{1}{2}\pi
_{na}^{*(m)}f_{bc}^{a}\eta _{m}^{c}\pi _{n}^{b}+  \nonumber \\
&&\frac{1}{12}\varepsilon _{npr}\pi _{na}^{*(m)}f_{bc}^{a}f_{de}^{c}\eta
_{m}^{e}\eta _{p}^{d}\eta _{r}^{b}-\frac{1}{2}\varepsilon _{mnp}\bar{A}%
_{a}^{(m)\alpha }f_{bc}^{a}\left( \left( D_{\alpha }\right) _{\;\;d}^{c}\eta
_{n}^{d}\right) \eta _{p}^{b}+  \nonumber \\
&&\lambda _{a}^{*(m)}\left( \frac{1}{2}f_{bc}^{a}\eta _{m}^{c}\lambda ^{b}-%
\frac{1}{12}\left( f_{bc}^{a}f_{de}^{c}+f_{dc}^{a}f_{be}^{c}\right) \eta
_{m}^{e}\eta _{n}^{d}\pi _{n}^{b}\right) +\frac{1}{6}\varepsilon _{mnp}%
\tilde{A}_{a}^{\alpha }f_{bc}^{a}f_{de}^{c}\left( \left( D_{\alpha }\right)
_{\;\;f}^{e}\eta _{m}^{f}\right) \eta _{n}^{d}\eta _{p}^{b}-  \nonumber \\
&&\frac{1}{2}\tilde{A}_{a}^{\alpha }f_{bc}^{a}\left( \left( \left( D_{\alpha
}\right) _{\;\;d}^{c}\eta _{m}^{d}\right) \pi _{m}^{b}-\eta _{m}^{c}\left(
D_{\alpha }\right) _{\;\;d}^{b}\pi _{m}^{d}\right) -\frac{1}{2}\bar{\eta}%
_{ma}^{(m)}f_{bc}^{a}\eta _{n}^{c}\pi _{n}^{b}+\bar{\eta}%
_{na}^{(m)}f_{bc}^{a}\eta _{n}^{c}\pi _{m}^{b}-  \nonumber \\
&&\frac{1}{6}\varepsilon _{mnp}\bar{\eta}_{ra}^{(m)}f_{bc}^{a}f_{de}^{c}\eta
_{r}^{e}\eta _{n}^{d}\eta _{p}^{b}+\frac{1}{2}\tilde{\eta}%
_{ma}f_{bc}^{a}\left( \frac{1}{12}\varepsilon _{npr}f_{de}^{c}f_{fg}^{e}\eta
_{m}^{g}\eta _{n}^{f}\eta _{p}^{d}\eta _{r}^{b}-\varepsilon _{mnp}\pi
_{n}^{c}\pi _{p}^{b}\right) +  \nonumber \\
&&\tilde{\eta}_{ma}f_{bc}^{a}\left( \eta _{m}^{c}\lambda ^{b}-\frac{1}{2}%
f_{de}^{c}\eta _{m}^{e}\eta _{n}^{b}\pi _{n}^{d}\right) +\frac{1}{2}%
\varepsilon _{mnp}\bar{\pi}_{na}^{(m)}f_{bc}^{a}\left( \eta _{p}^{c}\lambda
^{b}-\frac{1}{2}f_{de}^{c}\eta _{p}^{e}\eta _{r}^{b}\pi _{r}^{d}\right) + 
\nonumber \\
&&\frac{1}{2}\bar{\pi}_{na}^{(m)}\left( -f_{bc}^{a}\pi _{m}^{c}\pi _{n}^{b}+%
\frac{1}{12}\left( f_{bc}^{a}\left(
f_{de}^{c}f_{fg}^{e}+f_{fe}^{c}f_{dg}^{e}\right) +f_{gc}^{a}\left(
f_{de}^{c}f_{fb}^{e}+f_{fe}^{c}f_{db}^{e}\right) \right) \eta _{n}^{g}\eta
_{m}^{f}\eta _{p}^{d}\eta _{p}^{b}\right) +  \nonumber \\
&&\frac{1}{2}\bar{\lambda}_{a}^{(m)}\left( f_{bc}^{a}\left( \lambda ^{c}\pi
_{m}^{b}-\frac{1}{6}\varepsilon _{mnp}f_{de}^{c}\lambda ^{e}\eta
_{n}^{d}\eta _{p}^{b}\right) +\frac{1}{6}\left(
f_{bc}^{a}f_{de}^{c}+f_{dc}^{a}f_{be}^{c}\right) \pi _{m}^{e}\eta
_{n}^{d}\pi _{n}^{b}\right) +  \nonumber \\
&&\frac{1}{24}\bar{\lambda}_{a}^{(m)}\left( \varepsilon
_{mnp}f_{de}^{c}\left( f_{bc}^{a}f_{fg}^{e}+f_{fc}^{a}f_{bg}^{e}\right) \eta
_{n}^{g}\eta _{r}^{f}\eta _{p}^{d}\pi _{r}^{b}+\frac{1}{12}M_{bdfhi}^{a}\eta
_{n}^{i}\eta _{n}^{h}\eta _{m}^{f}\eta _{p}^{d}\eta _{p}^{b}\right) + 
\nonumber \\
&&\frac{1}{2}\tilde{\lambda}_{a}\left( f_{bc}^{a}\lambda ^{c}\lambda ^{b}-%
\frac{1}{3}\left( f_{bc}^{a}f_{de}^{c}+f_{dc}^{a}f_{be}^{c}\right) \lambda
^{e}\eta _{m}^{d}\pi _{m}^{b}+\frac{1}{6}f_{de}^{c}\left(
f_{bc}^{a}f_{fg}^{e}+f_{gc}^{a}f_{fb}^{e}\right) \eta _{m}^{g}\eta
_{n}^{f}\pi _{n}^{d}\pi _{m}^{b}\right) +  \nonumber \\
&&\left. \frac{1}{144}\varepsilon _{mnp}\tilde{\lambda}_{a}\left( \bar{M}%
_{bdfhi}^{a}\eta _{m}^{i}\eta _{m}^{h}\pi _{m}^{f}\eta _{n}^{d}\eta _{p}^{b}-%
\frac{1}{12}\tilde{M}_{bdfhjk}^{a}\eta _{m}^{k}\eta _{m}^{j}\eta
_{n}^{h}\eta _{n}^{f}\eta _{p}^{d}\eta _{p}^{b}\right) \right) \,,
\end{eqnarray}
where 
\begin{eqnarray}
\label{22}
&&M_{bdfhi}^{a}=\left( f_{ec}^{a}f_{bd}^{c}+f_{dc}^{a}f_{be}^{c}\right)
\left( f_{fg}^{e}f_{hi}^{g}+f_{hg}^{e}f_{fi}^{g}\right) +\left(
f_{ec}^{a}f_{bf}^{c}+f_{fc}^{a}f_{be}^{c}\right) \left(
f_{dg}^{e}f_{hi}^{g}+f_{hg}^{e}f_{di}^{g}\right) +  \nonumber \\
&&\left( f_{ec}^{a}f_{bh}^{c}+f_{hc}^{a}f_{be}^{c}\right) \left(
f_{fg}^{e}f_{di}^{g}+f_{dg}^{e}f_{fi}^{g}\right) \,,
\end{eqnarray}
\begin{equation}
\label{22a}
\bar{M}_{bdfhi}^{a}=f_{ce}^{a}f_{dg}^{c}\left(
f_{hf}^{e}f_{bi}^{g}-f_{if}^{e}f_{bh}^{g}+3f_{bf}^{e}f_{hi}^{g}\right)
+f_{cf}^{e}\left( f_{bg}^{c}\left(
f_{he}^{a}f_{di}^{g}-f_{ie}^{a}f_{dh}^{g}\right)
+3f_{be}^{a}f_{dg}^{c}f_{hi}^{g}\right) \,,
\end{equation}
\begin{eqnarray}
\label{22b}
&&\tilde{M}_{bdfhjk}^{a}=f_{be}^{c}\left(
f_{gc}^{a}f_{df}^{e}+f_{dc}^{a}f_{gf}^{e}\right) \left(
f_{hi}^{g}f_{kj}^{i}+f_{ki}^{g}f_{hj}^{i}\right) +  \nonumber \\
&&f_{de}^{c}\left( f_{gc}^{a}f_{kh}^{e}+f_{kc}^{a}f_{gh}^{e}\right)
\left( f_{bi}^{g}f_{fj}^{i}+f_{fi}^{g}f_{bj}^{i}\right) +
\nonumber \\
&&f_{de}^{c}\left( f_{gc}^{a}f_{fh}^{e}+
f_{fc}^{a}f_{gh}^{e}\right) \left(
f_{ki}^{g}f_{bj}^{i}+f_{bi}^{g}f_{kj}^{i}\right) \,.
\end{eqnarray}

Now, we develop a gauge-fixing procedure that ensures a direct equivalence
with the standard antibracket-antifield BRST formalism. To this end, we
begin by restoring an anticanonical structure for all the variables
(including the bar and tilde ones) in order to bring the classical master
equation of the Sp(3) BRST formalism to a more familiar form. We focus, for
example, on the first antibracket, and discard the other two. As we cannot
declare the existing variables excepting $\Phi ^{A}$ (original fields and
ghosts) and $\Phi _{A}^{*(1)}$ conjugated in the first antibracket, we need
to extend the algebra of the Sp(3) BRST tricomplex~\cite{sp3gen} by adding
the variables $\left( \rho _{2}^{A},\rho _{3}^{A},\kappa _{1}^{A},\mu
_{2}^{A},\mu _{3}^{A},\nu _{1}^{A}\right) $, which are respectively
conjugated in the first antibracket to $\left( \Phi _{A}^{*(3)},\Phi
_{A}^{*(2)},\bar{\Phi}_{A}^{(1)},\bar{\Phi}_{A}^{(3)},\bar{\Phi}_{A}^{(2)},%
\tilde{\Phi}_{A}\right) $. As explained in~\cite{sp3gen}, it is useful to
still add some more variables in order to realize a proper connection with
the gauge-fixing procedure from the standard antibracket-antifield
formalism. In view of this, we introduce the fermionic fields $\stackrel{%
(0,0,0)}{\varphi }^{a}$ that do not enter the original action, and hence are
purely gauge, endowed with the gauge invariances $\delta _{\xi }\varphi
^{a}=\xi ^{a}$. Consequently, we further enlarge the ghost sector with the
fields 
\begin{equation}
\label{sp3.120}
\left( \stackrel{(1,0,0)}{C}_{1}^{a},\stackrel{(0,1,0)}{C}_{2}^{a},\stackrel{%
(0,0,1)}{C}_{3}^{a},\stackrel{(0,1,1)}{p}_{1}^{a},\stackrel{(1,0,1)}{p}%
_{2}^{a},\stackrel{(1,1,0)}{p}_{3}^{a},\stackrel{(1,1,1)}{l}^{a}\right) \,,
\end{equation}
displaying the Grassmann parities $\varepsilon \left( C_{m}^{a}\right)
=\varepsilon \left( l^{a}\right) =0$, $\varepsilon \left( p_{m}^{a}\right)
=1 $. For notational simplicity, we make the collective notation $\varphi
^{I}=\left( \varphi ^{a},C_{m}^{a},p_{m}^{a},l^{a}\right) $. Thus, the
additional antifield spectrum will contain the variables $\left( \varphi
_{I}^{*(m)},\bar{\varphi}_{I}^{(m)},\tilde{\varphi}_{I}\right) ,\;m=1,2,3$.
As the sector corresponding to the new fields does not interfere in any
point with the original one, the solution to the master equation of the
Sp(3) BRST formalism associated with the overall gauge theory will be 
\begin{equation}
\label{sp3.132}
\bar{S}=S+\int d^{D}x\left( \varphi _{a}^{*(m)}C_{m}^{a}+\left( \varepsilon
_{mnp}C_{na}^{*(m)}+\bar{\varphi}_{a}^{(p)}\right) p_{p}^{a}-\left(
p_{ma}^{*(m)}+\bar{C}_{ma}^{(m)}-\tilde{\varphi}_{a}\right) l^{a}\right) \,.
\end{equation}
Now, we restore the anticanonical structure also with respect to the newly
added variables. We give up the second and third antibracket, and introduce
the variables $\left(
r_{2}^{I},r_{3}^{I},k_{1}^{I},m_{2}^{I},m_{3}^{I},n_{1}^{I}\right) $
respectively conjugated to $\left( \varphi _{I}^{*(3)},\varphi _{I}^{*(2)},%
\bar{\varphi}_{I}^{(1)},\bar{\varphi}_{I}^{(3)},\bar{\varphi}_{I}^{(2)},%
\tilde{\varphi}_{I}\right) $. Consequently, if $\bar{S}$ is solution to the
equation~(\ref{7}), then 
\begin{equation}
\label{23}
\bar{S}_{1}=\bar{S}+\int d^{D}x\left( \Phi _{A}^{*(2)}\mu _{2}^{A}+\Phi
_{A}^{*(3)}\mu _{3}^{A}+\bar{\Phi}_{A}^{(1)}\nu _{1}^{A}+\varphi
_{I}^{*(2)}m_{2}^{I}+\varphi _{I}^{*(3)}m_{3}^{I}+\bar{\varphi}%
_{I}^{(1)}n_{1}^{I}\right) \,,
\end{equation}
satisfies the equation $\left( \bar{S}_{1},\bar{S}_{1}\right) _{1}=0$, which
is precisely the standard classical master equation in the first antibracket.

With the solution~(\ref{23}) at hand, we can employ now the gauge-fixing
procedure from the standard BRST formalism. This means that we have to
choose a fermionic functional $\psi _{1}$, with the help of which we
eliminate half of the variables in favour of the other half. For
definiteness, we eliminate the variables
$\left( \Phi _{A}^{*(1)},\rho _{2}^{A},\rho
_{3}^{A},\kappa _{1}^{A},\bar{\Phi}_{A}^{(2)},\bar{\Phi}_{A}^{(3)},\tilde{%
\Phi}_{A}\right) $, together with $\left( \varphi
_{I}^{*(1)},r_{2}^{I},r_{3}^{I},k_{1}^{I},\bar{\varphi}_{I}^{(2)},\bar{%
\varphi}_{I}^{(3)},\tilde{\varphi}_{I}\right) $, and, in the meantime,
enforce the gauge-fixing conditions 
\begin{equation}
\label{sp3.134}
\rho _{2}^{A}=\rho _{3}^{A}=\kappa
_{1}^{A}=0\,, r_{2}^{I}=r_{3}^{I}=k_{1}^{I}=0\,,
\end{equation}
that can be implemented by taking $\psi _{1}=\psi _{1}\left[ \Phi ^{\Delta
},\mu _{2}^{\Delta },\mu _{3}^{\Delta },\nu _{1}^{\Delta }\right] $, where
we employed the notations $\Phi ^{\Delta }=\left( \Phi ^{A},\varphi
^{I}\right) $, $\mu _{2,3}^{\Delta }=\left( \mu
_{2,3}^{A},m_{2,3}^{I}\right) $ and $\nu _{1}^{\Delta }=\left( \nu
_{1}^{A},n_{1}^{I}\right) $. It is understood that a variable is eliminated
by one of the formulas $\mathrm{antifield}=\frac{\delta ^{L}\psi _{1}}{%
\delta \left( \mathrm{field}\right) }$,\ $\mathrm{field}=-\frac{\delta
^{L}\psi _{1}}{\delta \left( \mathrm{antifield}\right) }$, depending if it
is a `field' or an `antifield'. In this context, we emphasise that $\left(
\Phi ^{\Delta },\Phi _{\Delta }^{*(3)},\rho _{3}^{\Delta },\bar{\Phi}%
_{\Delta }^{(1)},\mu _{3}^{\Delta },\bar{\Phi}_{\Delta }^{(3)},\nu
_{1}^{\Delta }\right) $ are regarded as `fields', while $\left( \Phi
_{\Delta }^{*(1)},\rho _{2}^{\Delta },\Phi _{\Delta }^{*(2)},\kappa
_{1}^{\Delta },\bar{\Phi}_{\Delta }^{(2)},\mu _{2}^{\Delta },\tilde{\Phi}%
_{\Delta }\right) $ are viewed like their corresponding `antifields', where
we performed the obvious notations $\Phi _{\Delta }^{*(m)}=\left( \Phi
_{A}^{*(m)},\varphi _{I}^{*(m)}\right) $, $\bar{\Phi}_{\Delta }^{(m)}=\left( 
\bar{\Phi}_{A}^{(m)},\bar{\varphi}_{I}^{(m)}\right) $, $\kappa _{1}^{\Delta
}=\left( \kappa _{1}^{A},k_{1}^{I}\right) $, $\rho _{2,3}^{\Delta }=\left(
\rho _{2,3}^{A},r_{2,3}^{I}\right) $, and
$\tilde{\Phi}_{\Delta }=\left( \tilde{%
\Phi}_{A},\tilde{\varphi}_{I}\right) $. For a proper link with the
standard approach, we take 
\begin{equation}
\label{sp3.135}
\psi _{1}=\int d^{D}x\left( -\mu _{3a}^{(\varphi )}\partial ^{\alpha }\mu
_{2\alpha }^{a}+\left( \partial ^{\alpha }\mu _{3\alpha }^{a}\right) \mu
_{2a}^{(\varphi )}+\left( \partial ^{\alpha }A_{\alpha }^{a}\right) \nu
_{1a}^{(\varphi )}-\left( \partial ^{\alpha }\varphi _{a}\right) \nu
_{1\alpha }^{a}\right) \,,
\end{equation}
where we put an extra superscript between parentheses where necessary in
order to distinguish the variables of the same type that carry the same
indices. After performing the gauge-fixing process, and further eliminating
some auxiliary variables from the resulting action, we finally deduce the
gauge-fixed action 
\begin{eqnarray}
\label{28}
&&\bar{S}_{\psi _{1}}^{\prime }= \int d^{D}x\left( -\frac{1}{4}F_{\alpha
\beta }^{a}F_{a}^{\alpha \beta }+\left( \partial ^{\alpha }p_{ma}\right)
\left( D_{\alpha }\right) _{\;\;b}^{a}\eta _{m}^{b}+\left( \partial ^{\alpha
}C_{ma}\right) \left( D_{\alpha }\right) _{\;\;b}^{a}\pi _{m}^{b}+\right. 
\nonumber \\
&&\left( \partial ^{\alpha }\varphi _{a}\right) \left( - \left( D_{\alpha
}\right) _{\;\;b}^{a}\lambda ^{b}+\frac{1}{2}f_{bc}^{a}\left( \left( \left(
D_{\alpha }\right) _{\;\;d}^{c}\eta _{m}^{d}\right) \pi _{m}^{b}-\eta
_{m}^{c}\left( D_{\alpha }\right) _{\;\;d}^{b}\pi _{m}^{d}\right) \right) - 
\nonumber \\
&&\frac{1}{6}\varepsilon _{mnp}\left( \partial ^{\alpha } \varphi
_{a}\right) f_{bc}^{a}f_{de}^{c}\left( \left( D_{\alpha }\right)
_{\;\;f}^{e}\eta _{m}^{f}\right) \eta _{n}^{d}\eta _{p}^{b}-  \nonumber \\
&&\left. \frac{1}{2} \varepsilon _{mnp}\left( \partial ^{\alpha
}C_{ma}\right) f_{bc}^{a}\left( \left( D_{\alpha }\right) _{\;\;d}^{c}\eta
_{n}^{d}\right) \eta _{p}^{b}+l_{a}\left( \partial ^{\alpha }A_{\alpha
}^{a}\right) \right) \,.
\end{eqnarray}
The gauge-fixed action~(\ref{28}) can be checked to be invariant under the
gauge-fixed Sp(3) BRST transformations 
\begin{equation}
\label{29}
s_{m}A_{\alpha }^{a}=\left( D_{\alpha }\right) _{\;\;b}^{a}\eta
_{m}^{b}\,, s_{m}\varphi ^{a}=C_{m}^{a}\,,
\end{equation}
\begin{equation}
\label{30}
s_{m}\eta _{n}^{a}=\varepsilon _{mnr}\pi _{r}^{a}+\frac{1}{2}f_{bc}^{a}\eta
_{m}^{c}\eta _{n}^{b}\,, s_{m}C_{n}^{a}=\varepsilon _{mnr}p_{r}^{a}\,,
\end{equation}
\begin{equation}
\label{31}
s_{m}\pi _{n}^{a}=-\delta _{mn}\lambda ^{a}-\frac{1}{2}f_{bc}^{a}\eta
_{m}^{c}\pi _{n}^{b}+\frac{1}{12}\varepsilon _{npr}f_{bc}^{a}f_{de}^{c}\eta
_{m}^{e}\eta _{p}^{d}\eta _{r}^{b}\,, s_{m}p_{n}^{a}=
-\delta _{mn}l^{a}\,,
\end{equation}
\begin{equation}
\label{32}
s_{m}\lambda ^{a}=\frac{1}{2}f_{bc}^{a}\eta _{m}^{c}\lambda ^{b}-\frac{1}{12}%
\left( f_{bc}^{a}f_{de}^{c}+f_{dc}^{a}f_{be}^{c}\right) \eta _{m}^{e}\eta
_{n}^{d}\pi _{n}^{b}\,, s_{m}l^{a}=0\,.
\end{equation}
This completes the Lagrangian Sp(3) BRST approach to pure Yang-Mills theory.

To conclude with, in this paper we have constructed the Lagrangian Sp(3)
BRST symmetry for the Yang-Mills theory. Our procedure is based on the
triplication of the gauge transformations, and subsequently, on the
resolution of the extended classical master equation. With the solution of
this equation at hand, we develop a gauge-fixing procedure that leads to a
gauge-fixed action which is invariant under some gauge-fixed Sp(3) BRST
transformations.

\section*{Acknowledgment}
This work has been supported by a Romanian National Council for Academic
Scientific Research (CNCSIS) grant.

\end{document}